\pdfoutput=1
\documentclass[11pt]{article}

\usepackage[margin=1in]{geometry}
\usepackage{amsmath,amssymb}
\usepackage{graphicx}
\usepackage[numbers,sort&compress]{natbib} 
\usepackage{hyperref}
\usepackage{textcomp} 
\usepackage{authblk}  
\usepackage{caption} 

\title{An optically enhanced crystalline silicon allotrope: hydrogen passivated type II silicon clathrate}

\author[1]{Yinan Liu}
\author[1]{Joseph P. Briggs}
\author[2]{Sam Saiter}
\author[2]{Meenakshi Singh}
\author[1]{Carolyn A. Koh\thanks{Corresponding author: Carolyn A. Koh, Department of Chemical and Biological Engineering, Colorado School of Mines, Golden, CO 80401, USA. \texttt{ckoh@mines.edu}}}
\author[2]{P. Craig Taylor}
\author[2]{Michael Walker}
\author[3]{Khalid Mateen}
\author[3]{Moussa Kane}
\author[2]{Reuben T. Collins}

\affil[1]{Department of Chemical and Biological Engineering, Colorado School of Mines, Golden, CO 80401, USA}
\affil[2]{Department of Physics, Colorado School of Mines, Golden, CO 80401, USA}
\affil[3]{TotalEnergies E\&P Research and Technology, Houston, TX 77002, USA}

\date{} 

\begin{document}
\maketitle

\footnotetext{\textbf{Abbreviations:} PL, photoluminescence; TOF-SIMS, time-of-flight secondary ion mass spectrometry; EPR, electron paramagnetic resonance; XRD, X-ray diffraction; HF, hyperfine lines.}

\begin{abstract}
While Si clathrates have been explored as promising direct bandgap semiconductors, their practical optoelectronic performance has been limited by high doping levels and structural defects. Hydrogen has long been used to improve the optoelectronic quality of conventional Si, yet its role in clathrate structures remains unexplored. In this study, we demonstrate that hydrogen (deuterium) can be incorporated into type II Si clathrate framework using remote plasma treatment. This process leads to the formation of NaD and SiD complexes, which significantly reduce both the Na donor density and dangling bond defects. Electron paramagnetic resonance confirms nearly a tenfold decrease in Na-related donor states, resulting in the lowest doping level reported in Si clathrates to date. Following passivation, the integrated photoluminescence intensity increases by a factor of 40, accompanied by a blue shift of the main emission peak, consistent with a transition closer to the intrinsic band edge. A new emission peak at 930 nm, attributed to hydrogen-related recombination centers, also appears. These improvements remain stable up to $400\,^\circ\mathrm{C}$. Altogether, this work establishes hydrogen passivation as a viable strategy for enhancing light emission in Si clathrates and opens a new pathway toward their application in Si-based light-emitting diodes and other direct-bandgap optoelectronic devices.
\end{abstract}



\section{Introduction}\label{sec1}

Diamond-structured silicon (d-Si) has powered the microelectronics revolution, but its indirect bandgap remains a key limitation, preventing the efficient integration of electronics and photonics on a single chip, and requiring $\sim 100\,\mu\mathrm{m}$
 thick high-purity wafers for photovoltaic applications.\cite{soref2007past} Over the decades, various strategies have been explored in the search for an optically efficient Si-based material. For instance, amorphous silicon (a-Si), though widely used in thin-film photovoltaics, suffers from light-induced degradation.\cite{stutzmann1985role, stuckelberger2013comparison} Si-based superlattice structures offer tunable optical properties, but are hindered by complex synthesis procedures and difficulty achieving high quantum efficiencies.\cite{zheng2005present} Nanostructured approaches such as porous Si and Si quantum dots show promise, particularly for sensing and biomedical applications.\cite{leonardi2021biosensing} However, their integration into functional optoelectronic devices remains challenging due to the inherent trade-offs between quantum confinement and charge transport.\cite{garcia2021semiconductor} These limitations highlight the need for a crystalline Si material with strong optical absorption, good transport properties, and compatibility with device integration.\cite{wang2014direct, beekman2016clathrates, taylor2016exotic}

One of the most promising candidates is Si clathrates, which are cage-like, crystalline Si inclusion compounds. They are composed of tetrahedrally bonded Si atoms, like d-Si, but arranged into a framework of large and small cages.\cite{liu2025advancements} These open structures can be stabilized during synthesis by alkali metal guest atoms, typically Na, which occupy interstitial sites and act as shallow donors. \cite{schenken2020electron, ammar2004clathrate} This guest-host interaction enables the formation of the expanded clathrate framework under easily accessible growth conditions. In many ways, type II Si clathrates exhibit the ideal characteristics for optoelectronic applications: a nearly direct bandgap around 1.7 eV, strong optical absorption, and a crystalline structure favorable for electronic transport \cite{liu2021synthesis, beekman2006synthesis, bharwal2025enhancing} . In particular, type II Si clathrate with low Na content has an absorption coefficient about two orders of magnitude higher than d-Si, enabling the use of thin film absorbers.\cite{liu2021synthesis}  Photoluminescence (PL) studies reveal the material's ability to emit light at room temperature, underscoring its potential as a Si-based LED material.\cite{liu2021synthesis, bharwal2025enhancing, bharwal2024influence} Beyond optoelectronics, the recent spin dynamics investigations demonstrate that low-Na-content Si clathrates could have applications in quantum information science.\cite{briggs2024characterization} With these exceptional properties, Si clathrates are poised to impact next-generation semiconductor technologies. 

However, a critical gap remains between the exceptional properties of Si clathrates and their realization in practical devices. In type II Si clathrate, the  guest atoms, typically Na in $\mathrm{Na}_{x}\mathrm{Si}_{136}$ ($0 < x < 24$), essential for synthesis also act as donors, introducing high free carrier concentrations and hindering intrinsic semiconducting behavior.  Reducing the donor concentration is thus crucial for optoelectronic applications. Na can diffuse out of the clathrate cage structure at temperatures near $400\,^\circ\mathrm{C}$. Chemical approaches for Na removal, including acid etches (HF/HNO$_{3}$, HCl), vapor iodine treatment, and SF$_{6}$ reactive ion etching, have also been reported.\cite{liu2021synthesis, ammar2005preparation, krishna2014efficient, gryko2000low} Through these various chemical treatments and diffusion processes, Na concentrations have been reduced to the \(10^{18}\,\text{cm}^{-3}\) range, where the clathrate behaves as a heavily doped semiconductor.\cite{liu2021synthesis} However, the carrier concentration needs to be further reduced for effective device integration. Additionally, as in many d-Si or a-Si based devices, native recombination centers, such as dangling bond defects that can significantly degrade optical properties and device performance, are also observed in Si clathrates. \cite{schenken2020electron, yamaga2020electron} To date, almost all reports on electrical conductivity in type II Si clathrates indicate transport is affected by and in many cases dominated by disordered phases located near the grain boundaries,\cite{gryko2000low, mott1973properties, beekman2006synthesis} with little progress made in controlling these defects.

There have only been a few reports of devices fabricated from type II Si clathrate and these illustrate the doping and defect issues.   For instance, current-voltage (IV) characteristics of preliminary solar devices based on type II Si clathrates have failed to demonstrate typical "diode-like" behavior, instead showing characteristics consistent with tunneling and defect currents caused by high doping levels and defects in the material.\cite{fix2020silicon, kume2017group} Notably, surface photovoltage, a key indicator of solar cell potential, has been observed in Na-doped type II Si clathrates, but it varies significantly with Na concentration.\cite{bharwal2025enhancing, bharwal2024influence} This finding underscores the critical need to further minimize guest doping and defects to realize device-quality Si clathrate materials suitable for obtaining the intrinsic properties of Si clathrate and finally achieving semiconductor and photovoltaic applications. 

Hydrogen passivation is a versatile technique widely used to reduce defects and enhance semiconductor performance. In Si-based devices, it minimizes recombination losses, while in emerging materials like perovskites and compound semiconductors, it improves stability and electronic properties.\cite{lee2018review, zhang2023rationalization, johnson1986hydrogen} With hydrogen’s small size and ability to passivate unterminated bonds, it represents a novel and potentially significant alternative guest for enhancing the optoelectronic properties of Si clathrates, though its inclusion in type II Si clathrate has not been previously reported.

In this work, we used an inductively coupled plasma to incorporate the hydrogen isotope deuterium (D) into type II Si clathrate films and systematically explored how plasma conditions influence D incorporation, passivation effectiveness, and structural stability. Time-of-flight secondary ion mass spectrometry (TOF-SIMS) confirmed deuterium integration and the formation of NaD and SiD complexes. X-ray diffraction (XRD) verified that the metastable clathrate framework remains intact under optimized plasma conditions. Electron paramagnetic resonance (EPR) revealed that deuterium effectively passivates both Si dangling bonds and Na donor states, reducing the Na doping level to $\sim 4 \times 10^{17}~\mathrm{cm}^{-3}$, the lowest reported in Si clathrates. This substantial reduction in defects and carrier concentration led to a 40-fold enhancement in photoluminescence (PL) intensity and a blue shift toward the intrinsic band edge. These results provide the first direct evidence that hydrogen can enter the Si clathrate framework and effectively passivate defect states, significantly enhancing the optoelectronic performance. More broadly, this work establishes a new strategy for realizing high-quality, light-emitting Si materials, and lays the groundwork for extending hydrogen passivation to other passivating species and other direct bandgap Si allotropes for next-generation optoelectronic devices.

\section{Experimental Method}\label{sec2}

Deuterium-incorporated Si clathrate films were prepared by plasma treatment of Si clathrate films with exceptionally low Na content. The starting films were synthesized using a two-step method, as described in Ref.~\citenum{liu2021synthesis}, which we briefly summarize here. They were grown on $\langle 100 \rangle$ n-type Czochralski Si wafers doped with phosphorus ($5{-}10~\Omega\cdot\mathrm{cm}$). All processing steps were carried out either in an Ar-purged glovebox or in a tube furnace under flowing Ar or vacuum conditions.

The first step involves forming a NaSi film by diffusing Na into the Si substrate. To do this Na metal is placed in the bottom of a tantalum crucible with the Si wafer held directly above the Na. This assembly was annealed at $560\,^{\circ}\mathrm{C}$
 for 70 minutes in a quartz furnace tube under ultrahigh-purity Ar flow. The NaSi precursor film was then heated at $400\,^{\circ}\mathrm{C}$
 under vacuum for 24 hours in a quartz tube furnace where it decomposed into the Si clathrate phase. An additional 48-hour vacuum anneal at $430\,^{\circ}\mathrm{C}$
 was performed to further reduce excess Na. Subsequently, the films were rinsed with ethanol and deionized water to remove residual Na, followed by a 2-minute SF$_{6}$ dry etch (250 W, 13.56 MHz, 0.4 Torr) in a reactive ion etcher to remove a disordered surface layer and 
further lower the Na content. \cite{liu2021synthesis}

XRD has often been used to estimate Na content in the clathrate. Its sensitivity, however, decreases as $x$ approaches 1 in $\mathrm{Na}_{x}\mathrm{Si}_{136}$. Instead, we used EPR measurements of electron spin density as a more sensitive estimate of the Na donor level prior to and after the plasma treatment. Details of this procedure are given in Ref.~\citenum{schenken2020electron}. The spin density after the above synthesis and treatment was determined to be $2 \times 10^{18}~\mathrm{cm}^{-3}$, corresponding to $x \approx 0.009$ in $\mathrm{Na}_x\mathrm{Si}_{136}$. While this indicates that most large and small cages in the Si clathrate are guest-free, from the standpoint of electronic material applications, it remains a heavily doped semiconductor due to residual Na atoms, which act as shallow donors.

TOF-SIMS analysis discussed below revealed that as-prepared samples exhibited a low background of hydrogen prior to plasma treatment. The origin of this still needs investigation, but is likely due to an interaction of Na and trace water impurities introduced during synthesis. To differentiate between background hydrogen and that introduced during treatment, a deuterium (D) plasma was used. Deuterium plasma hydrogenation was conducted using remote downstream treatment, with the film placed within a quartz tube 27~cm from the plasma source. This placed the sample inside the heated zone of a tube furnace and the plasma source just outside of the furnace. The deuterium plasma was produced using a 13.56~MHz RF generator, with the flow rate maintained at 7~sccm and RF power adjusted between 100--250~W. Gas pressure was monitored by a capacitance manometer and varied from 150~mTorr to 800~mTorr. The plasma treatment duration ranged from 1 to 2~hours, during which the film was held at a fixed temperature between $300\,^\circ\mathrm{C}$ and $385\,^\circ\mathrm{C}$ in the surrounding furnace. To minimize deuterium out-diffusion at the end of the process, the region of the quartz tube containing the sample was rapidly quenched while the plasma remained on. The plasma was then turned off when the temperature dropped to approximately $70\,^\circ\mathrm{C}$.

The samples were characterized using XRD, Raman scattering spectroscopy, TOF-SIMS, PL, and EPR, with some of the techniques detailed in Ref.~\citenum{liu2021synthesis}. The ion intensities were acquired with an IonTOF TOF-SIMS system using a primary ion beam of Bi\textsuperscript{+} and a secondary ion beam of Cs\textsuperscript{+}. PL spectra were collected using an Acton 300i spectrometer with a 150~g/mm grating and a 100~µm entrance slit. A 550~nm long-pass filter was used to block scattered laser light from entering the spectrometer, resulting in the low-wavelength cutoff. An X-band Bruker EMX EPR instrument, equipped with a ColdEdge closed-cycle helium cryostat, was used to detect low-concentration Na guest species and alterations in the native defect states of the samples. EPR measurements were conducted across a temperature range of 4.6–300~K.

\begin{figure}[t]
\centering
\includegraphics{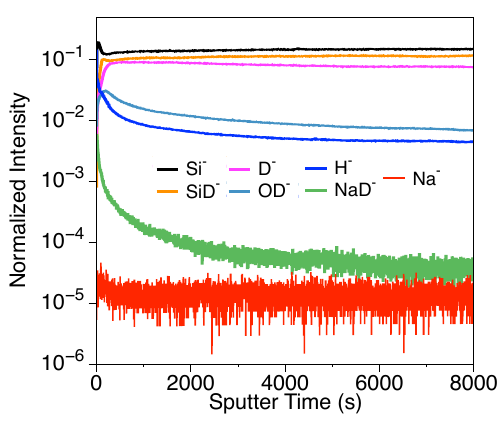}
\caption{Normalized TOF-SIMS ion intensities in low-Na type II Si clathrate after deuterium plasma treatment.}
\label{Fig1}
\end{figure}

\section{Results and Discussion}\label{sec3}

\subsection{Deuterium incorporation into Si clathrate film}

We successfully introduced hydrogen (deuterium) into low-Na-doped type II Si clathrate films for the first time using a remote RF plasma system. The sample shown in Fig. ~\ref{Fig1}  was treated at $385\,^\circ\mathrm{C}$ with 250~W RF power and 180~mTorr pressure for 2~hours. Prior to TOF-SIMS analysis, all samples were pumped overnight in an ultra-high vacuum environment to ensure vacuum equilibration and minimize surface contamination. This step was essential to reduce contributions from atmospheric H\textsubscript{2} or adsorbed water, allowing the detected deuterium-related signals to be confidently attributed to plasma incorporation rather than environmental sources. It is important to note that TOF-SIMS sensitivity varies substantially between elements. All of the measurements reported in this study used a Bi\textsuperscript{+} primary ion with an energy of 30~keV and a secondary Cs\textsuperscript{+} ion with an energy of 2~keV. Under Cs$^+$ sputtering, H$^-$ and Si$^-$ are detected with much higher sensitivity than Na$^-$.~\cite{wilson1991systematics} Therefore, the low Na intensity in Fig. ~\ref{Fig1} does not directly reflect the atomic concentration of Na relative to other species. To better assess the Na donor level, we rely on EPR measurements. The introduced deuterium formed both NaD and SiD complexes as shown in Fig. ~\ref{Fig1}. Since Na atoms are known to reside within the clathrate cages, the detection of NaD suggests that deuterium (and therefore H) can penetrate the cages, which is a critical requirement for effective passivation. The presence of SiD species suggests that deuterium also interacts with and passivates Si dangling bonds, which is confirmed by EPR as discussed below. Dangling bonds have previously been identified by EPR as the dominant intrinsic defect in type II Si clathrates.~\cite{schenken2020electron}

\begin{figure}[t]
\centering
\includegraphics{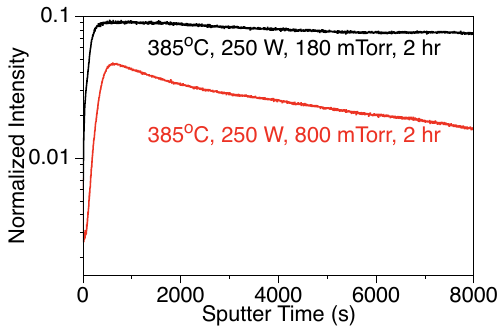}
\caption{Normalized TOF-SIMS deuterium ion intensities in low-Na type II Si clathrate films after plasma treatment under two different conditions.}
\label{Fig2}
\end{figure}

The intensities of deuterium-related complexes, such as NaD and SiD, were found to depend strongly on the plasma treatment conditions. To achieve optimal passivation, it is desirable to introduce a substantial amount of deuterium to form these compensating complexes, while preserving the integrity of the clathrate structure. In many reports on hydrogen passivation of crystalline Si (c-Si) and a-Si/c-Si heterojunctions,~\cite{rizk1991hydrogen, geissbuhler2013amorphous} high vacuum pressures, often above 700~mTorr, were used during RF plasma treatment. To assess the influence of pressure on deuterium incorporation, we compared TOF-SIMS results obtained under relatively high pressure (800~mTorr) and low pressure (180~mTorr), while keeping the temperature, power, and treatment duration constant. As shown in Fig. ~\ref{Fig2}, the deuterium incorporation level was significantly lower at 800~mTorr, likely due to a reduced fraction of atomic hydrogen or H$^+$ species under these conditions.\cite{abdel2006determination, mendez2006atom} These species, owing to their small size, are expected to be the primary contributors to hydrogen diffusion into the Si clathrate framework.

\begin{figure}[t]
\centering
\includegraphics{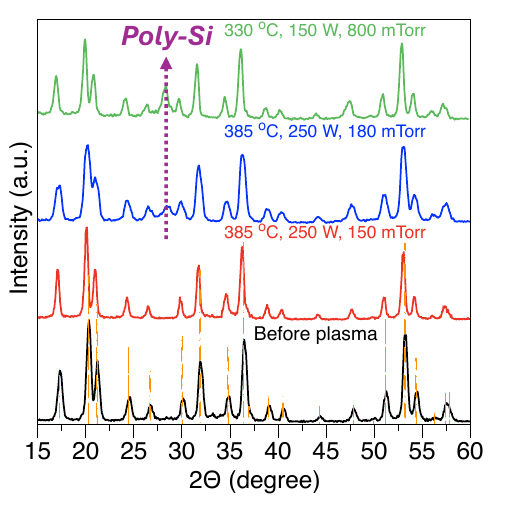}
\caption{XRD patterns of type II Si clathrate films before and after D$_2$ plasma treatment. Orange vertical lines indicate the reference pattern for type II Si clathrate (ICDD 98-024-8181). The purple dotted line marks the characteristic peak of poly-Si. All samples were treated for 2~hours under different plasma conditions.}
\label{Fig3}
\end{figure}

Stability of the metastable clathrate framework was also improved by using lower pressure. XRD measurements were performed before and after plasma treatments under various conditions to assess changes in the Si framework. At elevated temperatures ($\sim 400\,^\circ\mathrm{C}$), higher pressures ($\sim 800$~mTorr), or increased power levels, partial conversion to polycrystalline Si (poly-Si) was observed. The top XRD pattern in Fig. ~\ref{Fig3}, obtained at 800~mTorr (330$\,^\circ$C, 150~W), showed a higher poly-Si conversion despite lower temperature and power than the other traces in the figure, confirming that higher pressure increases structural degradation. Reducing the pressure to 180~mTorr resulted in only slight conversion to poly-Si, as shown in the blue pattern of Fig. ~\ref{Fig3}. In this case, XRD peaks also broadened compared to the pre-treatment pattern, suggesting increased structural inhomogeneity. However, when the pressure was further reduced to 150~mTorr while maintaining the same temperature and power, no signs of conversion to poly-Si were detected by XRD. These results highlight the importance of optimizing plasma conditions to enable effective hydrogen passivation without compromising the clathrate phase. Therefore, plasma treatments at 150--180~mTorr were used for most of the samples discussed in the following sections.

\subsection{Passivation of defects in hydrogenated Si clathrate film}

\begin{figure}[t]
\centering
\includegraphics{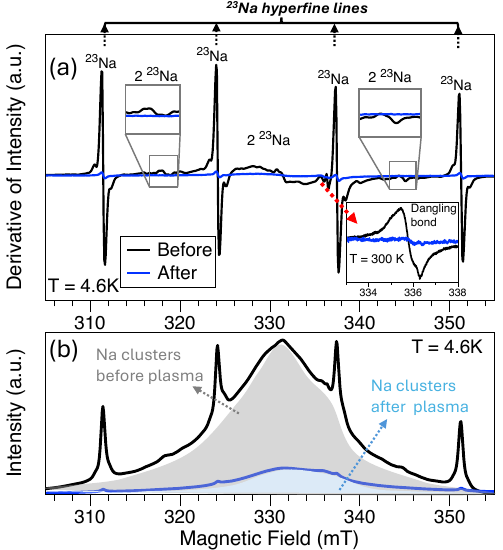}
\caption{(a) EPR spectra of low-Na type II Si clathrate at 4.6~K before and after deuterium plasma treatment; the inset shows room-temperature spectra. (b) Integral of the low-temperature spectra from (a). Both spectra were averaged over five scans, using a gain of $10^4$ and a 20~dB attenuation. The spectra before and after passivation were normalized by sample weight.}
\label{Fig4}
\end{figure}

 Fig. ~\ref{Fig4}a shows the low-temperature EPR spectra of a low-Na type II Si clathrate film before and after plasma treatment. The film was treated at $385\,^\circ\mathrm{C}$, 250~W, 180~mTorr for 2~hours. At this low Na concentration ($x \sim 0.009$ in $\mathrm{Na}_x\mathrm{Si}_{136}$), Na atoms are predominantly located inside the large cages, with their donor electrons localized at the nuclei at low temperatures.~\cite{schenken2020electron, krishna2014efficient} The dominant features and fine structure of the spectra have been previously identified and discussed in Refs.~\cite{schenken2020electron, ammar2004clathrate, yamaga2020electron}. Here, we focus on three key features: the four prominent $^{23}$Na hyperfine lines (HF) due to donor electrons on Na atoms ( Fig. ~\ref{Fig4}a); a broad peak in the integrated spectrum associated with clustered Na ( Fig. ~\ref{Fig4}b); and the dangling bond defect feature, which becomes more pronounced at room temperature, as shown in the inset of  Fig. ~\ref{Fig4}a. We note that the continuous wave (CW) EPR spectrum shown in  Fig. ~\ref{Fig4}a is obtained using a modulation technique, which yields the derivative of the absorption signal. The corresponding absorption spectrum, presented in  Fig. ~\ref{Fig4}b, is obtained by integrating the CW EPR spectrum. The low-temperature EPR spectra after passivation were recorded at both 20 dB and 25 dB attenuation. The slightly stronger signals observed at 20 dB attenuation suggest that the spectra were not significantly saturated by microwave power; however, minor saturation effects cannot be completely excluded.

After D$_{2}$ plasma treatment, all the Na-related EPR lines are significantly reduced in intensity ( Fig. ~\ref{Fig4}a and ~\ref{Fig4}b). We note that the Na did not leave, it remains in the cages under the plasma annealing conditions, as confirmed by TOF-SIMS profiles before and after plasma treatment.  Instead, the reductions in EPR line intensities are caused by hydrogen (deuterium) entering the cages and passivating the Na by forming NaD, consistent with TOF-SIMS results (Fig. ~\ref{Fig1}). Additionally, hydrogen (deuterium) also passivates the Si dangling bond defect, as evidenced by its decreased intensity in the inset of Fig. ~\ref{Fig4}a . This observation aligns well with previous studies on hydrogen passivation in a-Si.\cite{shi2016plasma} A range of samples with different Na and native defect concentrations were examined in EPR following deuterium passivation.  In all cases a significant reduction in the dangling bond and Na related features was observed.

An interesting insight arises from the similarity between the EPR spectral changes observed here and those we previously reported for Li-doped Si clathrates.\cite{liu2023formation} In previous studies, thermal diffusion of Li did not produce direct Li-related EPR signatures; instead, suppression of Na hyperfine lines and Na clusters was observed. This behavior was attributed to the formation of Li–Na complexes, which result in multiple occupancy in one cage and spin-singlet states that are EPR-silent.\cite{liu2023formation} Notably, both hydrogen and lithium belong to the same group in the periodic table, and are small enough to diffuse into the clathrate framework. In the present case, although TOF-SIMS revealed strong deuterium incorporation, no distinct hydrogen-related EPR signals were detected. This suggests that, similar to Li, the incorporated hydrogen (or deuterium) may form complexes within or on the clathrate cages, which are EPR-inactive. 

Overall, our findings demonstrate that incorporated hydrogen (or deuterium) interacts with interstitial Na guests and Si dangling bond defects, offering promising potential for mitigating both guest-induced doping and deep-level defects in clathrate materials. EPR spin density analysis indicates that the total Na doping level decreased from $2 \times 10^{18}~\mathrm{cm}^{-3}$ to $4 \times 10^{17}~\mathrm{cm}^{-3}$, nearly an order of magnitude reduction, which represents the lowest doping level reported in clathrates to date. The concentration of isolated Na (with no D pairing) alone was reduced by approximately 90\%. Additionally, the density of dangling bonds approached the background detection limit of the EPR instrument (inset of Fig. ~\ref{Fig4}a ). Following hydrogen/deuterium passivation, it is also important to investigate the thermal stability of this effect and determine up to what temperature the passivation can be preserved, which is critical for future device fabrication processes.

\subsection{Dehydrogenation of passivated Si clathrate films}

\begin{figure}[t]
\centering
\includegraphics[width=\linewidth]{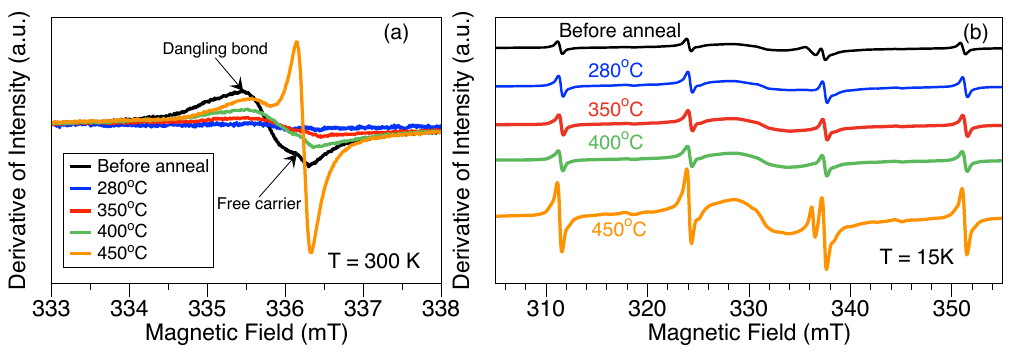}
\caption{(a) Room-temperature and (b) low-temperature EPR spectra of passivated Si clathrate films before and after a series of annealing steps in a sealed EPR tube. The films were sequentially annealed at $280\,^\circ\mathrm{C}$ for 7~hours, $350\,^\circ\mathrm{C}$ for 12~hours, $400\,^\circ\mathrm{C}$ for 12~hours, and $450\,^\circ\mathrm{C}$ for 12~hours. The low-temperature spectra were averaged over five scans using a gain of $10^4$ and 20~dB attenuation. The room-temperature spectra were averaged over ten scans using the same parameters.}
\label{Fig5}
\end{figure}

It has been reported that the hydrogen passivation effect in a-Si and at a-Si/c-Si interfaces degrades upon heating, as hydrogen gradually effuses from the a-Si:H layers.~\cite{Street1991} Similarly, in our study, we observed that the introduced deuterium leaves the material upon annealing. To better track the evolution of different EPR features during this process, we selected a sample with high initial defect density, as indicated by XRD, and performed plasma treatment at $385\,^\circ\mathrm{C}$, 250~W, 180~mTorr for 1~hour and 40~minutes. As shown in the black curve of Fig. ~\ref{Fig5}a , the passivated sample exhibits a pronounced dangling bond signal and a weak free carrier feature. After passivation, the sample was vacuum sealed in an EPR tube and subjected to four successive annealing steps at increasing temperatures, with the specific conditions provided in the Fig. ~\ref{Fig5} caption.

Fig. ~\ref{Fig5} shows the EPR spectra collected after each annealing step. The room-temperature data (Fig. ~\ref{Fig5}a) primarily reflect changes in the dangling bond signal, while the low-temperature spectra (Fig. ~\ref{Fig5}b) can help capture variations in the Na-related signal. After annealing at $280\,^\circ\mathrm{C}$, the dangling bond intensity decreased (blue curve), whereas the Na hyperfine lines showed a slight recovery. The binding energy required to remove a hydrogen atom from an ideal Si–H bond, creating a dangling bond, is approximately 3.55 eV,\cite{van1995silicon} which is slightly higher than the Na–H bond energy in NaH (2.2 eV).\cite{sawicka2003inverse} This suggests that deuterium released from NaD complexes may have been captured by Si dangling bonds via a deuterium exchange mechanism.

After annealing at higher temperatures ($350\,^\circ\mathrm{C}$ and $400\,^\circ\mathrm{C}$) for extended time (12 hours), the intensity of the Na HF lines remained nearly unchanged (Fig. ~\ref{Fig5}b), while the Si dangling bond signal showed a slight recovery, although still lower than the pre-anneal level (Fig. ~\ref{Fig5}a). This suggests that within this temperature range, deuterium may redistribute among different defect sites, but most of the passivating SiD and NaD complexes remain intact. At an even higher annealing temperature of $450\,^\circ\mathrm{C}$, where the Si clathrate phase remains stable, however, a more pronounced change was observed. The Na HF signal significantly increased, indicating that the majority of NaD complexes decompose only at this temperature. Similarly, the broad EPR background attributed to Na clusters also intensified at $450\,^\circ\mathrm{C}$ as shown in Fig. S1. Additionally, a sharper peak corresponding to Na free carriers, which is most evident in the room-temperature spectrum (yellow curve), grew noticeably, consistent with deuterium release from NaD and the reactivation of Na donors.

Based on the absence of deuterium-related EPR features, the ability of deuterium to obscure the Na EPR lines, and the recovery of these lines upon annealing, we speculate that D diffuses into the clathrate framework as D$^+$ or atomic D and then combines with Si dangling bonds and Na donors to passivate these defects. Some of these D$^+$ or D species may also recombine to form D$_2$, whose behavior is EPR silent. Fig.~S1 illustrates the changes in relative intensities of various EPR features, derived from the double integrals of the spectra shown in Fig. ~\ref{Fig5}. These results suggest that deuterium redistribution and possibly D$_2$ evolution occurs at temperatures below $400\,^\circ\mathrm{C}$, while most NaD complexes remain stable until annealing above $450\,^\circ\mathrm{C}$. Even after $450\,^\circ\mathrm{C}$, some deuterium may be still retained, as the Na HF signal does not fully recover to pre-passivation levels.

\subsection{Light emission properties}

\begin{figure}[t]
\centering
\includegraphics{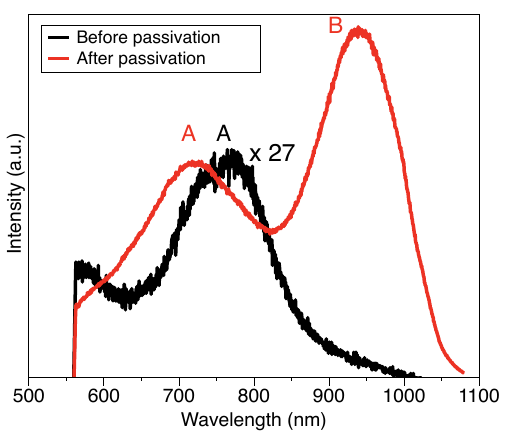}
\caption{Room-temperature PL spectra of the Si clathrate film before (black) and after (red) passivation. Both spectra were acquired using 10 s acquisition time.  Both spectra were background-corrected by subtracting the dark spectrum collected under the same conditions. The black curve was multiplied by a factor of 27 to allow comparison with the red curve on the same scale. We note that the shape of peak B beyond 1000 nm is affected by the detector cutoff. Also, the sharp cutoff at 560 nm is due to the long pass filter used to block the laser line.}
\label{Fig6}
\end{figure}

Hydrogen (deuterium) passivation significantly affects the PL emission properties of Si clathrate films. Fig. ~\ref{Fig6} compares PL spectra before and after passivation. Consistent with previously reported room-temperature PL spectra of type II Si clathrate films, a broad emission peak near 760~nm was observed before passivation ($\sim 1.63$~eV, labeled as peak A in the black curve of Fig. ~\ref{Fig6}). The same spectral shape and peak position were observed across the entire film surface, although the peak intensity varied slightly from spot to spot due to surface roughness and inhomogeneity. While the exact emission mechanism leading to peak A is not fully understood, as discussed in Ref.~\citenum{liu2021synthesis}, the peak is identified as arising from the main clathrate structure, and its emission energy aligns with the predicted bandgap of $\sim 2.0$~eV and the experimentally measured bandgap near $1.6\text{--}1.7~\mathrm{eV}$
.~\cite{gryko2000low, himeno2012optical, martinez2013synthesis, adams1994wide, demkov1994theoretical} We note that the slight rise in intensity at lower wavelength that may result from scattered residual laser light given the roughness of the sample, or from oxide-related emission, which has been reported but is not yet fully understood.~\cite{liu2021synthesis}

Also shown in Fig. ~\ref{Fig6} is the PL spectrum after passivation at $300\,^\circ\mathrm{C}$, 200~W, and 150~mTorr for 1~hour and 30~minutes. There are three dramatic changes in the spectrum. First, peak A shifts to shorter wavelength from 760~nm to 705~nm. Second, the emission intensity increases by almost a factor of 27. Finally, a new, rather strong emission band labeled peak B appears near 930~nm. These changes were consistently observed in other passivated films, although the size of the effects and the relative intensities of peaks A and B varied depending on the treatment conditions.

We speculate that the shift of peak A to shorter wavelength (higher energy, 1.76 eV) occurs because passivation suppresses some sub-bandgap recombination channels and results in emission closer to the intrinsic band edge.  Consistent with this, there have been reports of shifts in the energy of this peak with changes in Na concentration.\cite{bharwal2025enhancing} 

Perhaps more significant than the shift in peak A is the dramatic increase in its emission intensity after plasma treatment.   Fig. S2 presents a comparison of the integrated PL intensities of this peak before and after passivation. The integrated intensity increased by a factor of 40 after passivation, indicating a substantial reduction in nonradiative recombination. This enhancement is likely due to effective passivation of Si dangling bonds, Na-related defect states, and other non-radiative recombination centers, allowing a greater fraction of photoexcited carriers to recombine radiatively.

The emergence of peak B at 930~nm (1.33~eV) was observed after passivation. It was absent or too low to be detected in the untreated film. We do not believe this is due to the formation of any new crystalline or amorphous phases, as Raman spectra after passivation consistently show well-defined Si clathrate peaks across both the film surface and the buried interface, which was exposed by lifting off the film. Instead, this suggests the formation of hydrogen-related radiative centers. Interestingly, hydrogen-induced luminescence has been reported in hydrogenated amorphous silicon (a-Si:H) near 1.3--1.4~eV (885--950~nm) and attributed to localized states near the band edges.~\cite{dunstan1984photoluminescence, street1981luminescence, collins1980photoluminescence} In our Si clathrate films, the 930~nm emission is significantly suppressed under milder plasma conditions, such as lower pressure, power, temperature, and shorter treatment time, supporting its attribution to hydrogen-induced luminescent centers formed during passivation.

\subsection{Effect of thermal annealing of passivated samples on the PL spectrum}

\begin{figure}[t]
\centering
\includegraphics{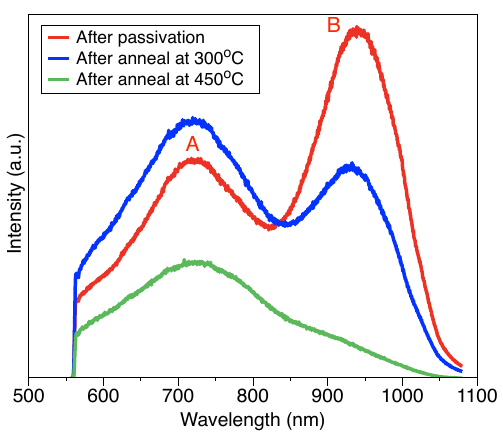}
\caption{Room-temperature PL spectra of the Si clathrate film after passivation and subsequent annealing steps. The red curve corresponds to the PL emission after passivation and is identical to the red curve shown in Fig. ~\ref{Fig6}. The film was then annealed under vacuum in a quartz tube at $300\,^\circ\mathrm{C}$ for 3~hours (blue curve), followed by a second anneal at $450\,^\circ\mathrm{C}$ for another 3~hours (green curve). All spectra were acquired under the same conditions and are plotted on the same scale.}
\label{Fig7}
\end{figure}

Interesting changes, which are consistent with the EPR results discussed above, were observed when the passivated samples were annealed under vacuum to promote the redistribution or out-diffusion of incorporated deuterium. As shown by the blue curve in Fig. ~\ref{Fig7}, annealing at $300\,^\circ\mathrm{C}$ for 3~hours led to a slight increase in the 705~nm peak and a concurrent decrease in the 930~nm peak. The enhanced 705~nm emission is consistent with EPR results, which showed a reduction of Si dangling bond defects after annealing a passivated sample at $280\,^\circ\mathrm{C}$ in a sealed EPR tube. We attribute this to thermally induced redistribution of deuterium atoms, which may passivate additional defect states that were not effectively reached during the initial plasma treatment.

When the sample was further annealed at $450\,^\circ\mathrm{C}$ for an additional 3~hours to drive out most of the incorporated deuterium, a noticeable reduction in the 705~nm peak was observed. This decrease correlates with the partial recovery of the dangling bond and Na-related lines in the EPR spectra, as shown in Fig.~S1 and Fig. ~\ref{Fig5}. Interestingly, the 930~nm peak became almost undetectable at room temperature. The suppression of this peak supports our hypothesis that it originates from hydrogen-related recombination centers introduced during plasma treatment, although the exact nature of these centers remains unclear. To quantify the evolution of the two peaks, we fitted the three PL spectra in Fig. ~\ref{Fig7} using two Gaussian components, then integrated their areas and plotted the normalized results in Fig.~S2. The 705~nm peak (peak A) was normalized to the intensity of peak A before passivation, while the 930~nm peak (peak B), which was nearly absent prior to passivation, was normalized to its integral in the green curve of Fig. ~\ref{Fig7}. As shown in Fig.~S2a, even after the $450\,^\circ\mathrm{C}$ anneal, the integrated intensity of peak A remained over 20 times higher than that of the unpassivated film. This is not surprising, given that the EPR signals associated with dangling bonds and Na hyperfine structures are not fully restored after the $450\,^\circ\mathrm{C}$ annealing, indicating that some deuterium remains trapped within the structure. Together, the PL and EPR results indicate that the passivation effects remain largely preserved up to approximately $400\,^\circ\mathrm{C}$, with significant degradation occurring only beyond this threshold.

\section{Conclusion}\label{sec5}

In this work, we demonstrate, for the first time, effective hydrogen (deuterium) passivation of low-Na-doped type II Si clathrate films using remote RF plasma treatment. Through a systematic combination of TOF-SIMS, EPR, XRD, and PL spectroscopy, we provide direct evidence that deuterium can penetrate the clathrate framework and form NaD and SiD complexes, significantly reducing both Na-induced donor states and Si dangling bond defects.

Spin density analysis from EPR indicates nearly an order-of-magnitude reduction in the Na donor level, reaching values as low as $\sim 4 \times 10^{17}~\mathrm{cm}^{-3}$, the lowest reported to date. This is particularly significant, as high Na donor levels have long been a major bottleneck in developing Si clathrates for optoelectronic applications. The density of Si dangling bonds also approached the instrumental detection limit after passivation. PL measurements revealed a 40-fold increase in integrated emission intensity and a notable blue shift from 760~nm to 705~nm, indicating suppression of sub-bandgap recombination and emission closer to intrinsic. Notably, a new emission peak near 930~nm was observed only after passivation, and its reduction after the thermal desorption suggests it originates from hydrogen-related recombination centers introduced during plasma treatment.

Importantly, annealing studies showed a strong passivation effect up to $\sim 400\,^\circ\mathrm{C}$
, indicating thermal stability compatible with common semiconductor processing conditions. This stability, combined with the dramatic enhancement in light emission, confirms that hydrogen (or deuterium) not only mitigates electrically and optically active defects, but also enables type II Si clathrate to function as an effective light-emitting material, advancing a long-standing goal in the development of Si-based optoelectronics.

These results open up promising opportunities for optoelectronic devices such as LEDs, photodetectors, and solar cells based on Si clathrates. Moreover, the demonstrated strategy may be extended to other low-density Si allotropes, including Si-Ge clathrates with tunable bandgaps \cite{baranowski2014synthesis, moriguchi2000first} and Si$_{24}$ with a direct gap near 1.5 eV,\cite{kim2015synthesis} offering a general approach for defect control and bandgap engineering in exotic group IV semiconductors. Overall, this work establishes hydrogen passivation as a powerful tool for unlocking the optoelectronic potential of Si clathrates and other clathrate-like materials.

\section*{Acknowledgments}
This work is supported by TotalEnergies EP Research and Technology. J.B., S.S., and M.S. acknowledge the support from the National Science Foundation under Grant No.~2114569. TOF-SIMS measurements were performed at the Colorado School of Mines using equipment supported by the National Science Foundation under Grant No.~1726898.

\section*{Conflict of interest}
The authors declare no potential conflict of interests.

\bibliographystyle{unsrtnat}
\bibliography{wileyNJD-AMA}

\clearpage
\renewcommand{\thefigure}{S\arabic{figure}}
\setcounter{figure}{0}

\section*{Supplementary Materials}

\captionsetup{justification=justified,singlelinecheck=false}

\begin{center}
\includegraphics[width=0.80\textwidth]{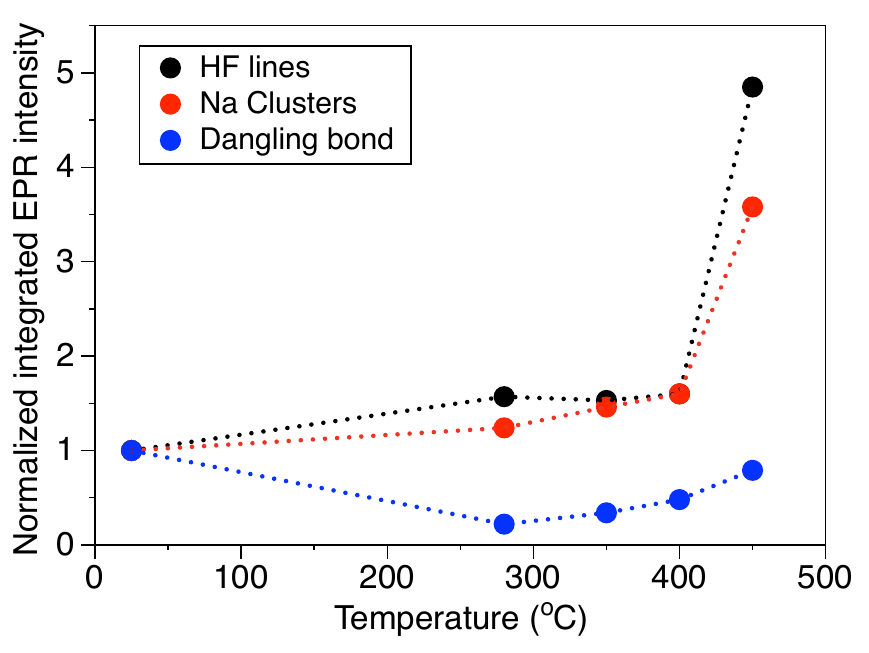}
\captionof{figure}{Evolution of relative EPR feature intensities during sequential annealing of a low-Na type II Si clathrate film passivated by remote RF plasma (250~W, 180~mTorr, $385\,^\circ\mathrm{C}$, 1~h 40~min). The passivated sample was sealed in a quartz EPR tube under vacuum and annealed at $280\,^\circ\mathrm{C}$ (7~h), $350\,^\circ\mathrm{C}$ (12~h), $400\,^\circ\mathrm{C}$ (12~h), and $450\,^\circ\mathrm{C}$ (12~h). EPR spectra were collected at room temperature and 15~K before annealing and after each annealing step using identical acquisition parameters. The low-temperature spectra were averaged over five scans using a gain of $10^4$ and 20~dB attenuation, while the room-temperature spectra were averaged over ten scans using the same settings. The plots show how the relative intensities of selected EPR features change with annealing temperature. These features include Na hyperfine (HF) lines (black), Na clusters (red), and dangling bonds (blue). Intensities were obtained from the double integrals of the EPR spectra in Fig.~\ref{Fig5}b and normalized to their respective values before annealing, as represented by the black curve in Fig.~\ref{Fig5}b. This analysis highlights the thermal response of different spin centers and provides insight into the stability of the passivation effect.}
\label{FigS1}
\end{center}

\vspace{1.5em}

\begin{center}
\includegraphics[width=0.80\textwidth]{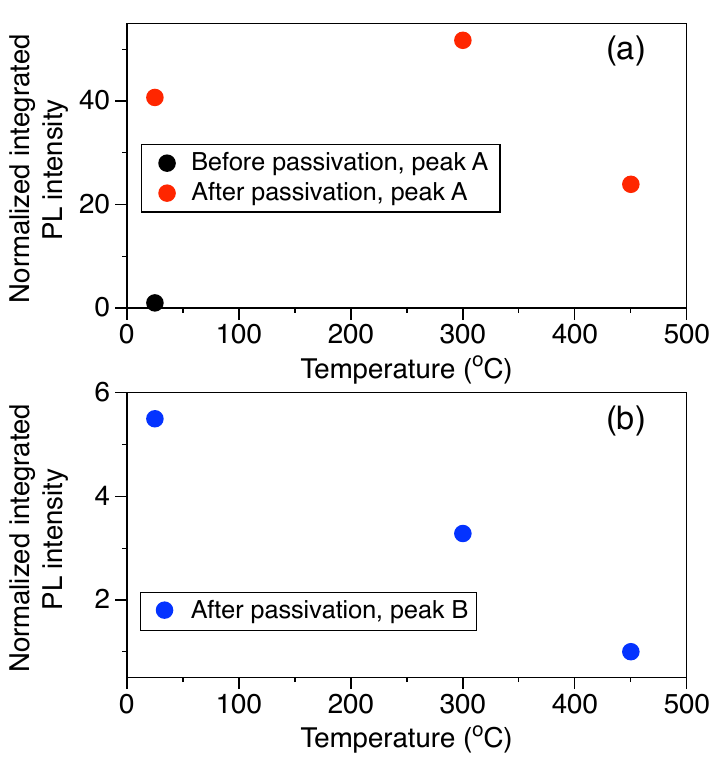}
\captionof{figure}{Integrated photoluminescence (PL) peak intensities for a low-Na type II Si clathrate film that was passivated using remote RF plasma (200~W, 150~mTorr, $300\,^\circ\mathrm{C}$, 1~h 30~min) and subsequently annealed under vacuum in a quartz tube at $300\,^\circ\mathrm{C}$ (3~h) and $450\,^\circ\mathrm{C}$ (3~h). PL spectra were collected at room temperature after passivation and after each annealing step under identical acquisition conditions as shown in Fig.~\ref{Fig7}. The spectra were fitted with two Gaussian components to extract the integrated intensities of the main PL peaks. Panel~(a) displays the evolution of peak~A (705~nm) before passivation (black dot) and after each anneal (red dots), normalized to the pre-passivation value. Panel~(b) shows the corresponding trend for peak~B (930~nm) (blue dots), which was not detectable prior to passivation and became prominent afterward, normalized to the value measured after the $450\,^\circ\mathrm{C}$ anneal. This figure highlights the thermal evolution of radiative centers and the stability of hydrogen-induced emission features.}
\label{FigS2}
\end{center}

\end{document}